\documentclass[12pt,twoside,pointlessnumbers,smallheadings]{revtex4}
\usepackage{amsmath}
\usepackage{amssymb}
\usepackage{color}

\usepackage{axodraw4j}

\usepackage{hangcaption,fancybox,feynmp}
\usepackage{indentfirst}
\usepackage{graphicx}
\usepackage{epsfig,subfigure,psfrag}
\usepackage{CJK}
\usepackage{mathrsfs}

\begin{document}


\title{
New Color-Octet Vector Boson Revisit

}
\author{Xiao-Ping Wang $^{1}$, You-Kai Wang $^{1}$, Bo
Xiao$^{1}$, Jia Xu $^{1}$ and Shou-hua
Zhu$^{1,2}$}

\affiliation{ $ ^1$ Institute of Theoretical Physics $\&$ State Key
Laboratory of Nuclear Physics and Technology, Peking University,
Beijing 100871, China \\
$ ^2$ Center for High Energy Physics, Peking University, Beijing
100871, China }

\date{\today}

\maketitle

\begin{center}

{\bf Abstract}

\begin{minipage}{15cm}
{\small  \hskip 0.25cm

Motivated by CDF recent measurements on di-jet invariant mass spectrum where di-jet is associated production with charged leptons ($e/\mu$) and missing
energy, we re-examine the previous proposed massive color-octet axial-vector-like boson $Z_c$. Our simulation showed that the di-jet bump around 120-160 GeV can be induced by
$Z_c$ with effective coupling $g_{Z_c q \bar q} =0.2 g_s$ (q represents the quark other than top and $g_s$ is the strong coupling constant). Moreover our numerical investigation indicated that the top quark forward-backward asymmetry
$A_{FB}^t$ can be reproduced without distorting shape of differential cross section $d\sigma/d M_{t\bar t}$, provided that the $Z_c$ and top quark coupling is appropriately chosen
($g_{Z_c t \bar t} \simeq 4.5 g_{Z_c q \bar q}$). Our results also showed that the theoretical $A_{FB}^t$ as functions of $\Delta y$ and $M_{t\bar t}$ can be consistent with
data within $1\sigma$ and $1.8\sigma$ respectively.

}

\end{minipage}
\end{center}


\newpage

\section{Introduction\label{introduction}}

In the previous work \cite{Xiao:2010ph}, part of authors proposed a new massive color-octet vector boson $Z_c$ just above $2 m_t$ in order to account for the
top quark forward-backward asymmetry $A_{FB}^t$ at Tevatron. Just after \cite{Xiao:2010ph} posted, a new analysis on $A_{FB}^t$ appeared \cite{Aaltonen:2011kc}. The analysis
indicated that, in the $t\bar t$ rest frame,  $A_{FB}^t$ increase with the $t\bar t$ rapidity difference $\Delta y$, and with the invariant mass
$M_{t\bar t}$ of the $t\bar t$ system. In order to satisfy the measured $A_{FB}^t$ for $M_{t\bar t}< 450$ GeV which is consistent with standard model (SM)
prediction, $m_{Z_c}$ must be adjusted very carefully to ensure the cancelation of contributions to asymmetric cross section. This feature makes the idea less
attractive. However the alternative proposals to account for anomalous  $A_{FB}^t$, for example the t-channel flavor changing $Z^\prime/W^\prime$ contribution and a generic s-channel heavy color octet, will have a risk to distort the shape of $d\sigma/d M_{t\bar t}$, especially for the high energy regime. This disadvantage is, in fact, one of our primary motivations to introduce the light color-octet axial-like vector boson.

Recently CDF at Tevatron released their measurements on WW/WZ cross sections via the
electron/muon + missing energy + di-jet channel \cite{Aaltonen:2009vh,CDFemu2jetlatest,cdflatest}.
The cross sections are in agreement with the SM prediction. However there seems an unexpected bump in the di-jet
spectrum around 120-160 GeV, though the significance ($3.2 \sigma$ \cite{cdflatest}) is not significant. The instant investigations showed that
the extra vector bosons \cite{Wang:2011uq,Cheung:2011zt}, as well as the new resonance in Technicolor models \cite{Eichten:2011sh}, can account for such bump with appropriate
parameters. However the new particles prefer to couple with quarks instead of leptons provided that the severe constraints from other measurements at LEP, Tevatron and even the
UA2 experiments. Such feature motivates us to pursue the possibility that the new bump is actually the color-octet axial vector. The obvious reason is that the color-octet does not
couple with leptons.

The interaction Lagrangian of $Z_c$ and quarks can be written as
\begin{eqnarray}
\mathscr{L} =i \overline{q} \left( g_V+ g_A \gamma_5\right) \gamma_\mu  T^a q Z_{c}^{\mu,a} +h.c.
\end{eqnarray}
and $g_V/g_A$ are vector- and axial-vector couplings among quarks and $Z_c$.
In order to ensure the successful prediction on $d\sigma/d M_{t\bar t}$ in the SM, the extra contribution to the total cross section should be limited.
Thus for simplicity we choose $g_V=0$, same with the choice in Ref. \cite{Xiao:2010ph}. This point can be understood from the amplitude squared for
the process $q\bar q \rightarrow t\bar t$
 \begin{equation} \begin{array}{rl} \sum\limits_{Color, Spin}|M|^2=& \frac{C_A C_F}{2}\{4g_s^4 (1+c^2+4m^2)\\\\
&+\frac{8g_s^2
\hat{s}(\hat{s}-M_G^2)}{(\hat{s}-M_G^2)^2+M_G^2\Gamma_G^2} [g_V^q g_V^t(1+c^2+4m^2)+2g_A^q g_A^t c] +\frac{4\hat{s}^2}{(\hat{s}-M_G^2)^2+M_G^2\Gamma_G^2}\times \\\\
&
[((g_V^q)^2+(g_A^q)^2)\times((g_V^t)^2(1+c^2+4m^2)+(g_A^t)^2(1+c^2-4m^2))
+8 g_V^q g_A^q g_V^t g_A^t c]\},
\end{array}
\label{axigluon}
\end{equation}
where $m=m_t/\sqrt{\hat{s}}$, $\beta=\sqrt{1-4m^2}$,$c=\beta
\cos\theta$ and  $g_V^{q (t)}/g_A^{q (t)}$ are vector- and axial-vector couplings among light quarks (top) and $Z_c$.
Here the terms at rhs represent QCD amplitude squared, interference between QCD and $Z_c$ amplitudes and $Z_c$ amplitude squared respectively.

Why can such light particle $Z_c$ escape the constraints from (1) new resonance search using di-jet invariant mass distributions; (2) quark composite scale limits which are derived
from the deviation from the background (mainly from QCD processes) shape of $P_T^j$ or $m_{jj}$?
The reasons are as following.
\begin{enumerate}

\item Jet is usually not measured so well as that of charged leptons and the di-jet mass peak can be buried by large mass resolution if the $Z_c$ contribution is
less than the huge QCD backgrounds. As the consequence, the Tevatron constraints on new particle less than 200 GeV are quite weak \cite{Buckley:2011vc}. The
UA2 experiment can only effectively constrain such light particle with the coupling strength larger than $\mathscr{O} (g_s)$. The required parameter to account
for the new bump is only  $0.2 g_s$, as shown below, which is permitted by direct searches of past experiments.

\item The shapes and magnitudes of $P_T^j$ and $m_{jj}$ after including the $Z_c$ contributions are not distorted severely, especially at high energy region (much larger than $m_{Z_c}$). As such the quark composite scale limit is {\em not} applicable to $Z_c$.

\end{enumerate}

The paper is organized as following. Section II showed that $W+Z_c$ associated production
can account for the di-jet invariant mass distribution observed by CDF \cite{cdflatest}. In section III  we presented the results for $A_{FB}^t$ and compared with measurements by CDF \cite{Aaltonen:2011kc}.
Section IV contained our conclusions and discussions.

\section{Di-jet associated production with $e/\mu$ and missing energy at Tevatron}

\begin{figure}[htbp]
\begin{center}
\scalebox{0.5}{\fcolorbox{white}{white}{
 \begin{picture}(276,304) (175,-75)
    \SetWidth{1.0}
    \SetColor{Black}
    \Line[arrow,arrowpos=0.5,arrowlength=5,arrowwidth=2,arrowinset=0.2](176,192)(256,128)
    \Line[arrow,arrowpos=0.5,arrowlength=5,arrowwidth=2,arrowinset=0.2](256,128)(256,32)
    \Line[arrow,arrowpos=0.5,arrowlength=5,arrowwidth=2,arrowinset=0.2](256,32)(176,0)
    \Photon(256,128)(336,144){7.5}{4}
    \Photon(256,32)(320,16){7.5}{3}
    \Line[arrow,arrowpos=0.5,arrowlength=5,arrowwidth=2,arrowinset=0.2](320,16)(416,32)
    \Line[arrow,arrowpos=0.5,arrowlength=5,arrowwidth=2,arrowinset=0.2](368,-48)(320,16)
    \Line[arrow,arrowpos=0.5,arrowlength=5,arrowwidth=2,arrowinset=0.2](336,144)(384,208)
    \Line[arrow,arrowpos=0.5,arrowlength=5,arrowwidth=2,arrowinset=0.2](400,112)(336,144)
    \Text(208,208)[lb]{\Large{\Black{$q$}}}
    \Text(176,32)[lb]{\Large{\Black{$\overline{q}^\prime$}}}
    \Text(416,0)[lb]{\Large{\Black{$q$}}}
    \Text(384,-80)[lb]{\Large{\Black{$\overline{q}$}}}
    \Text(288,160)[lb]{\Large{\Black{$W$}}}
    \Text(304,48)[lb]{\Large{\Black{$Z_c$}}}
    \Text(400,192)[lb]{\Large{\Black{$e/\mu$}}}
    \Text(416,128)[lb]{\Large{\Black{$\nu$}}}
  \end{picture}
}}
\end{center}

\caption{\label{fdww} Feynman diagram for $Z_c$ production at Tevatron.
}

\end{figure}
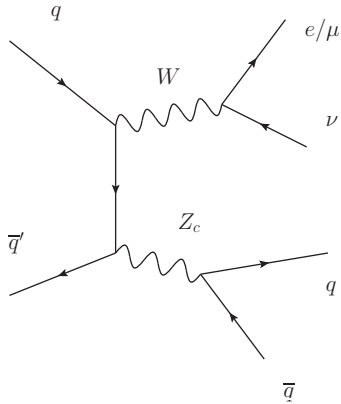

$Z_c$ and $W$ associated production is via
\begin{eqnarray}
p\overline{p} \rightarrow q \overline{q}^\prime \rightarrow W (\rightarrow e/\mu+\nu)+ Z_c (\rightarrow q\overline{q}),
\end{eqnarray}
and the Feynman diagram is shown in Fig. \ref{fdww}.

\begin{figure}[htbp]

\begin{center}
\includegraphics[width=0.80\textwidth]
{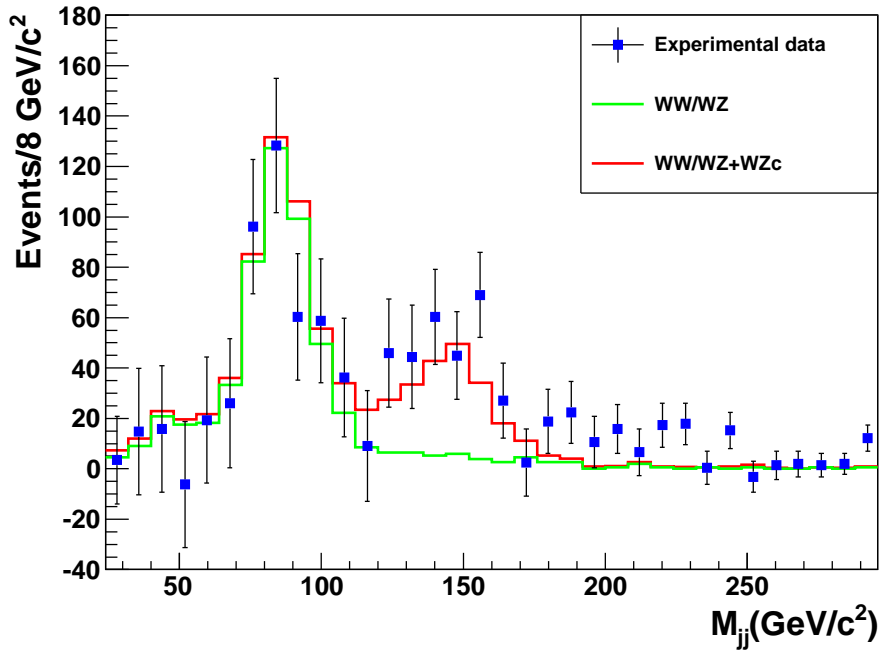}
\end{center}

\caption{\label{mjjatcdf} Di-jet invariant mass distribution at Tevatron with integrated luminosity $\mathscr{L}=4.3~fb^{-1}$. Data is taken from Ref. \cite{cdflatest}.
}

\end{figure}

In order to account for the CDF data, we simulate the di-jet invariant distribution arising from $WW/WZ$ as well as $W Z_c$ production with $m_{Z_c}=140$ GeV.
We choose the benchmark parameter set as $g_V=0$ and $g_A=0.2\ g_s$ as discussed above, which do not contradict with other measurements. The di-jet invariant distribution is shown
in Fig. \ref{mjjatcdf} after imposing the same cuts with those of \cite{cdflatest}. The events are generated
by MadGraph \cite{Maltoni:2002qb}, then the initial state radiation, final state radiation and fragmentation are carried out by Pythia \cite{Sjostrand:2006za}. The detector response is simulated by PGS. We corrected the jet energy according to Ref. \cite{phdthesis}.
From the Fig. \ref{mjjatcdf}, we can see that the new contribution arising from
the extra $Z_c$ can excellently fit the data.  Generally speaking, the new contributions from $Z_c$ mainly depends on the magnitude of $g_V - g_A$. For specific choice of {\em right-handed} coupling, namely
 $g_V - g_A=0$, the cross section is zero if parton mass is neglected.
 Such kind of $Z_c$ signal is not difficult to be examined at current running LHC, similar to the case of deci-weak $W^\prime/Z^\prime$ \cite{Wang:2011uq}.

\section{ Top Quark Forward-backward Asymmetry $A_{FB}^t$ at Tevatron}

Now that we know roughly the coupling among $Z_c$ and quarks, we switch to discuss whether the same parameters can account for
top quark forward-backward asymmetry $A_{FB}^t$ at Tevatron.  Since the choice of $g_V=0$ affects $d\sigma/dM_{t\bar t}$ insignificantly, we
don't show the numerical results here. Instead we will focus
on the $A_{FB}^t$ as functions of $\Delta y$ and $M_{t\bar t}$.

\begin{figure}[htbp]

\begin{center}
\includegraphics[width=0.80\textwidth]
{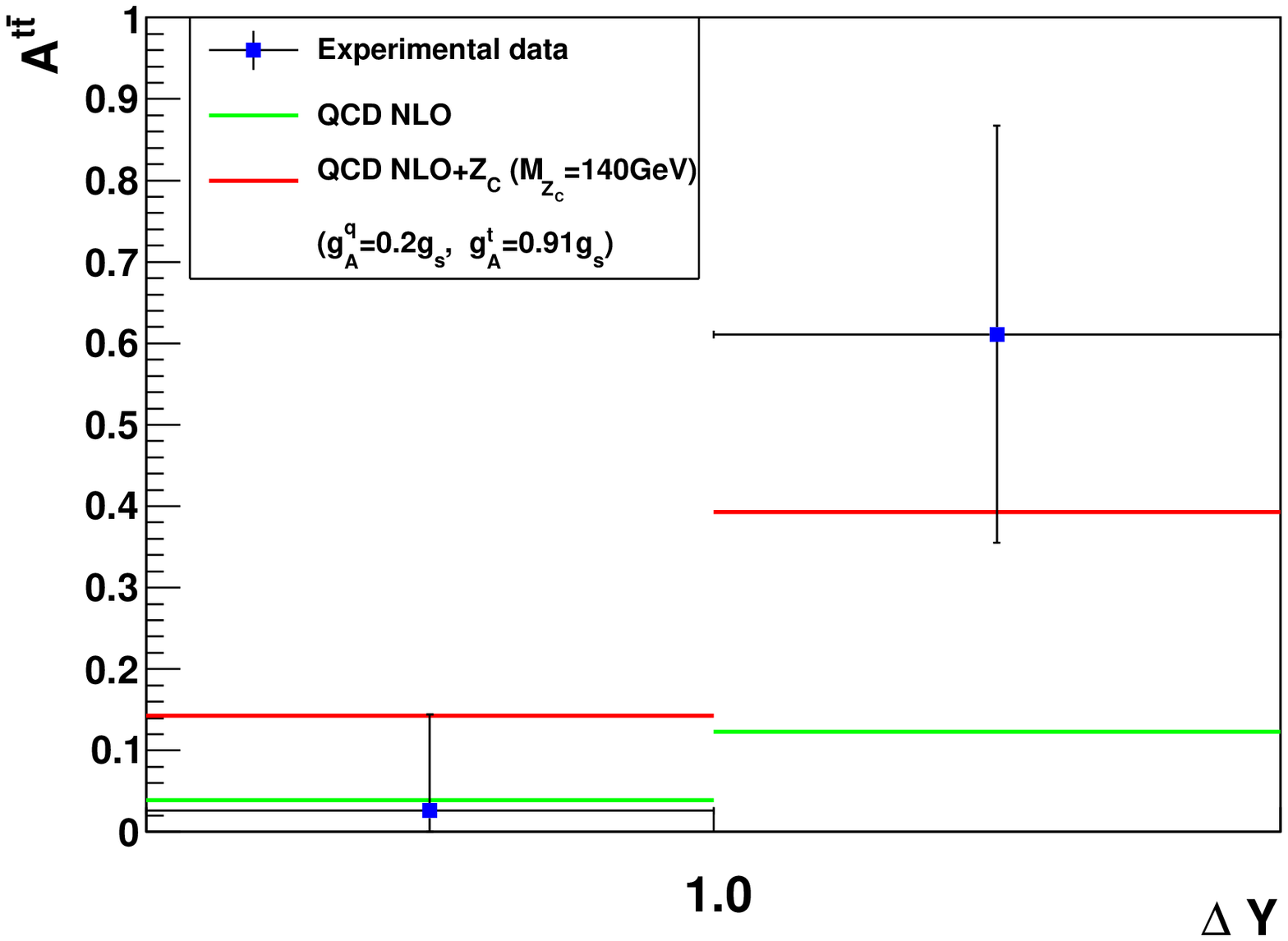}
\end{center}

\caption{\label{afby} $A_{FB}$ as a function of $\Delta y$ at Tevatron. Data is taken from Ref. \cite{Aaltonen:2011kc}.
}

\end{figure}

\begin{figure}[htbp]

\begin{center}
\includegraphics[width=0.80\textwidth]
{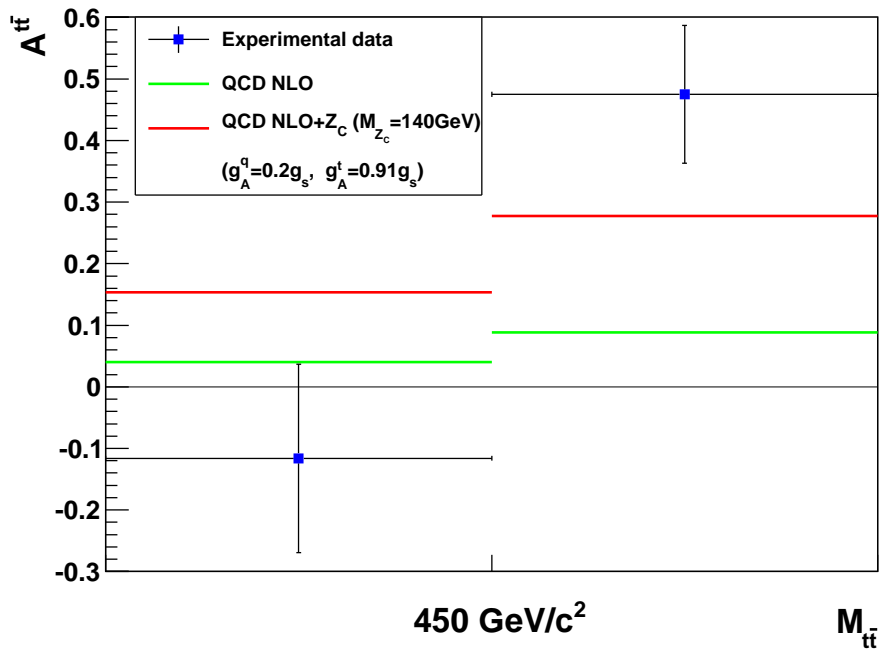}
\end{center}

\caption{\label{afbmtt} $A_{FB}$ as a function of $M_{t\bar t}$ at Tevatron. Data is taken from Ref. \cite{Aaltonen:2011kc}.
}

\end{figure}

The numerical investigation indicated that if we choose $g_A^t=g_A^q$, we can't obtain the observed  $A_{FB}^t$. Instead if we choose
$$
g_A^t/g_A^q=4.5,
$$
$A_{FB}^t\simeq 20\%$. For comparison the experimental measurement is $A_{FB}^t\simeq 15.8\% \pm 7.5\%$. In Fig. \ref{afby} and
\ref{afbmtt}, we also present $A_{FB}^t$ as functions of $\Delta y$ and $M_{t\bar t}$. From the figures, we can see that  data and
theory is in agreement within $1\sigma$ for $\Delta y$ distribution while about $1.8 \sigma$ for $M_{t\bar t}$ distribution. It is obvious that
the $Z_c$ plus SM contributions can improve the agreement between theory and data greatly.

\section{Conclusions and discussions }

Motivated by recent measurement of di-jet distribution associated with $e/\mu$ and missing energy by CDF, we re-examine the
previous idea, namely the new massive color-octet axial-vector-like boson $Z_c$, to account for $A_{FB}^t$ observation. Our
numerical results showed that $Z_c$ can explain the new di-jet bump within allowed parameters. We also investigated whether such
$Z_c$ can account for $A_{FB}^t$. We found that if $g_A^t/g_A^q \simeq 4.5$ the theory and data is in excellent agreement. Moreover
the $A_{FB}^t$ distributions as functions of $\Delta y$ and $M_{t\bar t}$ are consistent with data within $1\sigma$ and about $1.8\sigma$ respectively.

At the LHC, the new $Z_c$ can be easily discovered via the $W Z_c$ associated production. Once $Z_c$ is discovered the detailed properties, such as
spin, coupling structure etc., can also be studied in top pair production processes. Here the observables of one-side forward-backward asymmetry \cite{Wang:2010du} and/or edge charge asymmetry \cite{Xiao:2011kp} can be
utilized to analyze the event samples.

\section*{Acknowledgment}

We would like to thank Jia Liu for the stimulating discussion. This work was supported in part by the Natural Sciences Foundation
of China (No 11075003).


\begin{thebibliography}{10}
\bibitem{Xiao:2010ph}
  B.~Xiao, Y.~k.~Wang and S.~h.~Zhu,
  arXiv:1011.0152 [hep-ph].

\bibitem{Aaltonen:2011kc}
  T.~Aaltonen {\it et al.}  [CDF Collaboration],
  arXiv:1101.0034 [hep-ex].


\bibitem{Aaltonen:2009vh}
  T.~Aaltonen {\it et al.}  [CDF Collaboration],
  Phys.\ Rev.\ Lett.\  {\bf 104}, 101801 (2010)
  [arXiv:0911.4449 [hep-ex]].

\bibitem{CDFemu2jetlatest}
  {\em http://www-cdf.fnal.gov/physics/ewk/2010/WW\_WZ/index.html.}


\bibitem{cdflatest}
  T.~Aaltonen {\it et al.}  [CDF Collaboration],
  arXiv:1104.0699 [hep-ex].



\bibitem{Wang:2011uq}
  X.~P.~Wang, Y.~K.~Wang, B.~Xiao, J.~Xu and S.~h.~Zhu,
  arXiv:1104.1161 [hep-ph].

\bibitem{Cheung:2011zt}
  K.~Cheung and J.~Song,
  arXiv:1104.1375 [hep-ph].


\bibitem{Eichten:2011sh}
  E.~J.~Eichten, K.~Lane and A.~Martin,
  arXiv:1104.0976 [hep-ph].



\bibitem{Buckley:2011vc}
  M.~R.~Buckley, D.~Hooper, J.~Kopp and E.~Neil,
  arXiv:1103.6035 [hep-ph].












\bibitem{Maltoni:2002qb}
  F.~Maltoni and T.~Stelzer,
  JHEP {\bf 0302}, 027 (2003)
  [arXiv:hep-ph/0208156].



\bibitem{Sjostrand:2006za}
  T.~Sjostrand, S.~Mrenna and P.~Z.~Skands,
  JHEP {\bf 0605}, 026 (2006)
  [arXiv:hep-ph/0603175].

\bibitem{phdthesis} V. Cavaliere (2010), Fermilab Ph.D Thesis 2010-51,\\
  http://www.slac.stanford.edu/spires/find/hep/www?r=FERMILAB-THESIS-2010051.


\bibitem{Wang:2010du}
  Y.~k.~Wang, B.~Xiao and S.~h.~Zhu,
  Phys.\ Rev.\  D {\bf 82}, 094011 (2010)
  [arXiv:1008.2685 [hep-ph]];
  Y.~k.~Wang, B.~Xiao and S.~h.~Zhu,
  Phys.\ Rev.\  D {\bf 83}, 015002 (2011)
  [arXiv:1011.1428 [hep-ph]].




\bibitem{Xiao:2011kp}
  B.~Xiao, Y.~K.~Wang, Z.~Q.~Zhou and S.~h.~Zhu,
  Phys.\ Rev.\  D {\bf 83}, 057503 (2011)
  [arXiv:1101.2507 [hep-ph]].



\end{thebibliography}
\end{document}